\documentclass[prl,showpacs,twocolumn,amsmath,amssymb,superscriptaddress]{revtex4}

\usepackage{graphicx}% Include figure files
\usepackage{dcolumn}% Align table columns on decimal point
\usepackage{natbib}
\usepackage{bm}% bold math
\usepackage{color}

\newcommand{\Sr}{Sr$_3$Cr$_2$O$_8$}

\begin{document}                  % DO NOT DELETE THIS LINE
\bibliographystyle{apsrev}

\title{Orbital ordering promotes weakly-interacting S=1/2 dimers in the triangular lattice compound Sr$_3$Cr$_2$O$_8$ }

\author{L. C. Chapon}
 \email{L.C.Chapon@rl.ac.uk}
 \affiliation{ISIS, Rutherford Appleton Laboratory - STFC, OX11 0QX, United Kingdom}
\author{C. Stock}
 \affiliation{ISIS, Rutherford Appleton Laboratory - STFC, OX11 0QX, United Kingdom}
\author{P. G. Radaelli}
 \affiliation{ISIS, Rutherford Appleton Laboratory - STFC, OX11 0QX, United Kingdom}
\author{C. Martin}
 \affiliation{Laboratoire CRISMAT-UMR, 6508 ENSI CAEN, 6 Marechal Juin, 14050, France} 

\date{\today}

\begin{abstract}
The weakly interacting S=1/2 dimers system \Sr\ has been investigated by powder neutron diffraction and inelastic neutron scattering. Our data reveal a structural phase transition below room temperature corresponding to an antiferro-orbital ordering with nearly 90 $^{\circ}$ arrangement of the occupied \textit{3z$^2$-r$^2$} d-orbital. This configuration leads to a drastic reduction of the inter-dimer exchange energies with respect to the high temperature orbital-disorder state, as shown by a spin-dimer analysis of the super-superexchange interactions performed using the Extended Huckel Tight Binding method. Inelastic neutron scattering reveals the presence of a quasi non-dispersive magnetic excitation at 5.4 meV, in agreement with the picture of weakly-interacting dimers.        
\end{abstract}


\maketitle                        % DO NOT DELETE THIS LINE
The dimerization of a magnetic S=1/2 system, namely the formation of a spin-singlet ground state at low temperature with opening of a gap ($\Delta$) in the excitation spectrum, has been observed to date in a variety of inorganic crystals. In the limit of weak inter-dimer interactions, the problem maps exactly into the physics of Bose-Einstein Condensates (BEC), as explained in a recent review \cite{ISI:000254559900011} and references therein. Upon application of an external magnetic field above a critical value (H$_c$), one can excite triplons into the network of S=0 dimers, a direct analogy to increasing the density of Bosons in a condensate, here tunable by the strength of the magnetic field. This effect has been directly observed in prototypical systems of weakly interacting dimers such as TlCuCl$_3$ \cite{ISI:000087653300045} and BaCuSi$_2$O$_6$ \cite{ISI:000223472400064}. Recent work has been directed towards more complex systems, such as SrCu$_2$(BO$_3$)$_2$, where the presence of geometrically frustrated nearest-neighbor interactions strongly increases the repulsion of magnons, promoting new types of ground states. \\
\indent Very recently, a new family of dimerized antiferromagnet, namely A$_3$M$_2$O$_8$ (A=Ba,Sr M=Mn,Cr), has shown BEC of magnons (H$_c$=12.5 Tesla for Ba$_3$Cr$_2$O$_8$ \cite{Ba3Cr2O8APS} and 9.2T for Ba$_3$Mn$_2$O$_8$ \cite{ISI:000234335400103}). A unique feature in this series is the existence of isostructural compounds with variant spin state (S=1/2 for Cr and S=1 for Mn). The systems also uniquely show the presence of dimerized MO$_4^{3-}$ tetrahedra with an M ion in the unusual 5+ oxidation state and the presence of competing exchange interactions, since the dimers are arranged in a triangular lattice presenting a high degree of geometrical frustration. In all A$_3$M$_2$O$_8$ compounds, the dimers, aligned along the hexagonal \textit{c}-axis, are equivalents by symmetry and characterized by an intra-dimer interaction J$_0$ (following the convention of \cite{ISI:000169623600050}). Each dimer has three nearest neighbors in the adjacent layer (interaction J$_1$), six next-nearest neighbors in-plane (interaction J$_2$) \cite{ISI:000169623600050} and six further neighbors (interactions J$_3$) in adjacent planes \cite{ISI:000242899400048}. A striking feature is the variation of the effective inter-dimer effective exchange energy J’ (defined as the sum of all inter-dimer interactions J'=3J$_1$+6J$_2$+6J$_3$) between Mn (J’/J$_0$=1.4) and Cr compounds (J’/J$_0$=0.3 for Ba and J’/J$_0$=0.1 for Sr). In fact, a recent inelastic study by Stone et al. has revealed that Ba$_3$Mn$_2$O$_8$, magnetically gapped with $\Delta$=11.2K, is far from the non-interacting limit since it displays strongly dispersive magnetic modes and long distance competing interactions. Although one might expect some variation due to chemical pressure, nothing in the known crystal structures justifies a reduction of J’/J$_0$ by a factor of 14 between Ba$_3$Mn$_2$O$_8$ and Sr$_3$Cr$_2$O$_8$.\\
\indent In the present letter, we show that the weak inter-dimer interactions in Sr$_3$Cr$_2$O$_8$ originate from a peculiar orbital ordering super-structure pre-formed at high temperature, involving nearly 90 degrees arrangement of the occupied \textit{3z$^2$-r$^2$} chromium d-orbital. By employing neutron diffraction, more sensitive than X-ray when probing the oxygen positions, we discovered a cooperative Jahn-Teller distortion corresponding to an axial compression along one of the CrO$_4^{3-}$ tetrahedron edge. This distortion lifts the electronic degeneracy (doublet) of the chromium-d electronic states, stabilizing the \textit{3z$^2$-r$^2$} level lying lower in energy than the x$^2$-$y^2$. A spin-dimer analysis on the basis of the Extented Huckel Tight Binding (EHTB) method, shows a large renormalization of the intra and inter-dimer energies upon orbital ordering, stabilizing a system of weakly-interacting dimers. The ratio J'/J$_0$=0.11 estimated by this method reproduces accurately the values previously reported from bulk measurements \cite{ISI:000248487900015}. Unlike for Ba$_3$Mn$_2$O$_8$\cite{stone:237201,ISI:000255457200059}, inelastic neutron scattering on polycrystalline sample of \Sr\ confirms that a weakly-dispersive gap is formed with energy of 5.4 meV. \\           
\indent The polycrystalline samples were prepared by solid state reaction in air. SrCO$_3$ and Cr$_2$O$_3$ were weighted in the 3:1 stochiometric ratio, carefully mixed and crushed. The powder was then pressed in the form of pellets, heated at 1200 $^{\circ}$ C during 3 days followed by quenching to room temperature.
Powder neutron diffraction data were collected on the General Materials Diffractometer (GEM) at the ISIS facility of the Rutherford Appleton Laboratory, UK. A 1.2g polycrystalline sample, was enclosed in a thin-walled cylindrical vanadium can placed in an Oxford Instrument Cryostat. Data have been recorded on warming from 1.6K to 295K. Rietveld analysis have been carried out with the program FullProf \cite{FullProf}. Inelastic Neutron Scattering experiments were performed on the MARI direct chopper spectrometer (ISIS). A 14 g sample was placed in a thin foil Al container, wrapped cylindrically in annular geometry. A Gd fermi chopper spun at 200 Hz was used to produce an incident neutron beam of 25 meV.  A disc chopper was used in conjunction with the fermi chopper to remove high-energy neutrons.  
\begin{figure}[h!]
\includegraphics[scale=0.30]{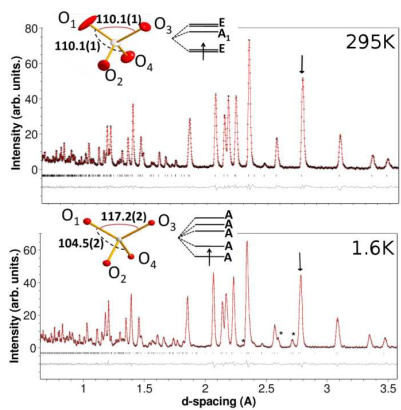}
\caption{ (Color Online) Rietveld refinements of neutron powder diffraction patterns at 295K and 1.6K for \Sr\ . The cross symbols and (red) solid lines represent the experimental and calculated intensities, respectively while the thin grey lines represent the difference between observed and calculated intensities. Black tick marks indicate the positions of Bragg peaks (see text for details). The inset shows the corresponding distortions in the CrO$_4^{3-}$ tetrahedra and associated splitting of the Cr-d levels.}
\label{Fig:Rietveld}
\end{figure}
\indent All the diffraction peaks at 295K can be indexed with the rhombohedral space group \textit{R$\overline{3}$m}, as previously proposed \cite{xxx} for this class of compounds. The Rietveld refinement, shown in Fig. \ref{Fig:Rietveld}, is of excellent quality (agreement factors R$_{F^2}$=2.3 \%\, $\chi^2$=2.0). However, reproducing accurately the Bragg intensities requires an unrealistically large anisotropic displacement parameter (ADP) on one of oxygen site (labelled O$_1$ following ref. \cite{xxx} ), as shown in the inset of Fig. \ref{Fig:Rietveld}. This ion corresponds to the apical oxygen in the bi-tetrahedral dimer unit. Furthermore, this ADP increases on cooling, confirming an anomalous behavior. Below 275K, several superlattice reflections are observed in the diffraction pattern, indicating a structural phase transition. All superlattice peaks can be indexed by a propagation vector \textbf{k}=(0,0,$\frac{3}{2}$) with respect to the R-centred hexagonal setting (\textbf{T} point of the Brillouin zone), equivalent to \textbf{k}=($\frac{1}{2}$,$\frac{1}{2}$,$\frac{1}{2}$) in the rhombohedral setting. Selected superlattice peaks are marked by an asterisk, in the bottom panel of Fig. \ref{Fig:Rietveld}. Moreover, some of the fundamental peaks become splitted below 275K, as best evidenced on the 1,1,0 reflection marked by an arrow in Fig. \ref{Fig:Rietveld}, indicating an additional monoclinic distortion. Four isotropy subgroups \cite{isotropy} of the high temperature (HT) trigonal space group can account for the observed symmetry lowering, with unit-cells all related to the parent cell by the transformation a$_m$=a$_h$-b$_h$,b$_m$=-a$_h$-b$_h$,c$_m$=$\frac{1}{3}$a$_h$-$\frac{1}{3}$b$_h$+$\frac{2}{3}$c$_h$ where the lattice vectors for the hexagonal and monoclinic cells are represented by the h and m subscripts.\\               
\begin{table}[!h]
\begin{tabular}{llllll}
\hline
\textbf{295K} & & & & & \\
\textit{Site(W)} & x & y & z & B(\AA\ $^2$) & \\
\hline
Sr$_1$(3a) & 0 & 0 & 0 & 1.17(6) &  \\
Sr$_2$(6c) & 0 & 0 & 0.20326(8) & 0.73(4) &  \\
Cr(6c) & 0 & 0 & 0.4058(1) & 0.39(7) &  \\
O$_1$(6c) & 0 & 0 & 0.3234(9) & 2.73(7) & \\
O$_2$(18h) & 0.8321(1) & 0.1679(1) & 0.89852(5) & 1.02(4) &  \\
\textbf{J$_0$=1150} & \textbf{J$_1$=177} & \textbf{J$_2$=178} & \textbf{J$_3$=1} & &   \\
\textbf{J'/J$_0$=1.4} & & & & & \\
\hline
\textbf{1.6K} & & & & & \\
\textit{Site(W)} & x & y & z & B(\AA\ $^2$) & \\
\hline
Sr$_1$(4e) & 0 & 0.2850(5) & $\frac{1}{4}$ & 0.14(4) & \\
Sr$_2$(8f) & 0.1002(4) & 0.2479(3) & 0.5545(1) & 0.17(2) & \\
Cr(8f) & 0.2020(7) & 0.2544(7)  & 0.8582(2) & 0.22(5) & \\
O$_1$(8f) & 0.1581(4) & 0.3129(4) & 0.7351(1) & 0.35(3) & \\
O$_2$(8f) & 0.1161(4)& 0.7800(5) & 0.5996(3) & 0.25(3) & \\
O$_3$(8f) & 0.8545(3) & -0.0104(5) & 0.6038(2) & 0.25(3) & \\
O$_4$(8f) & 0.3744(3)& 0.9858(5) & 0.5894(2) & 0.25(3) & \\
\textbf{J$_0$=2530} & \textbf{J$_1'$=59} & \textbf{J$_1''$=4} & \textbf{J$_2'$=4} & &   \\
\textbf{J$_2''$=47} & \textbf{J$_3'$=7} & \textbf{J$_3''$=0} & \textbf{J'/J$_0$=0.11} & & \\ 
\hline
\end{tabular}
\caption{ Structural parameters and values of the $ <\left( \Delta E \right)^2>$ energies in meV$^2$, calculated for various exchange
paths at 295K and 1.6K using the EHTB method (see text for details). The atomic positions of all sites are presented with related Wyckof letters (W) and isotropic thermal parameters. At 295K, the data have been refined with space group R$\overline{3}$m (lattice parameters a=5.57218(5) $\AA$ , c=20.1703(3) $\AA$). At 1.6K, the space group is C2/c, unique-axis b, setting 1 of \cite{IUCR} (lattice parameters a=9.6638(2) $\AA$ , b=5.5437(1) $\AA$ , c=13.7882(2) $\AA$ , $\beta$ =103.666(2) $^{\circ}$)  }
\label{Table:crystallo}
\end{table} 
Out of the four isotropy subgroups, two solutions with space groups \textit{C2/c} satisfy the observed systematic absences of Bragg peaks. Only one model, corresponding to a single mode of the T3- bidimensional representation \cite{isotropy} allows to reproduce the observed intensities. The result of the Rietveld refinement corresponding to this model for data at 1.6K is shown in the bottom panel of Fig. \ref{Fig:Rietveld}. The refinement is of excellent quality with agreement factors R$_{F^2}$ of 2.3\%\ and $\chi^2$=1.7, making the low temperature (LT) structure unambiguous. The crystallographic parameters extracted for the high temperature and low temperature phases are reported in Table \ref{Table:crystallo}.
\begin{figure}[!h]
\includegraphics[scale=0.30,angle=-90]{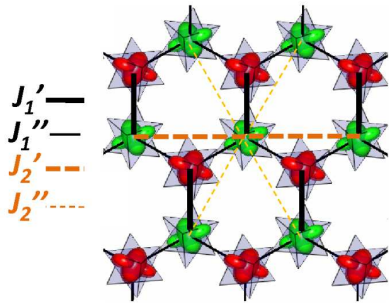}
\caption{ (Color Online) Orbitally ordered super-structure in the low temperature phase of \Sr\ . The structure is projected along the pseudo 3-fold axis, lost in the transition. The tetrahedral CrO$_4^{3-}$ units are shown as light grey polyhedra. The 3z$^2$-r$^2$ orbitals are sketched with different colors depending on their direction (see text for details). The various inter-dimer exchange interactions are marked by different connecting lines.}
\label{Fig:OO}
\end{figure}
\indent The primary order parameter of the transition is the antiferrodistortive displacement of the O$_1$ ion. This is coupled to the displacement of the Sr$_1$ ions, positioned in the same layer as O$_1$ ions, and a slight rotation of the O$_2$ ions in the tetrahedral basal plane, which are splitted into three orbits in the LT structure (O$_{2,3,4}$). The main feature of the transition, presented in the inset of Fig. \ref{Fig:Rietveld}, is the large modification of some of the tetrahedral angles, in particular O$_1$-Cr-O$_3$ and O$_1$-Cr-O$_4$, respectively 117.2(1)$^{\circ}$ and 104.5(1)$^{\circ}$, deviating substantially from the values found in an ideal tetrahedron (109.9$^{\circ}$). In the HT trigonal structure, the point symmetry of the Cr site is 3m (C$_{3v}$ in Schoenflies notation), i.e. the crystalline electric field splits the d orbitals of the Cr ion into non-bonding E orbitals and antibonding A$_1$+E orbitals. The single electron of Cr$^{5+}$ occupies the degenerate low-lying E state, formed by the \textit{3z$^2$-r$^2$} and \textit{x$^2$-y$^2$} orbitals. In the LT structure, where all point symmetry elements of the Cr site are lost, the E level split into 2 sublevels. The compression along the O$_1$-Cr-O$_3$ edge of the tetrahedron favors a low-lying {3z$^2$-r$^2$} level, similarly to an axial compression along one of the $\overline{4}$-axis (S$_4$ in Schoenflies notation) of a perfect tetrahedron. Here the 3z$^2$-r$^2$ orbital is directed towards the middle of the tetrahedron edge formed by the O$_1$ and O$_3$ ions. Since all CrO$_4^{3-}$ tetrahedra in the LT structure (eight per unit cell) are equivalent by symmetry, one can derive indirectly the orbital ordering pattern simply from the positions of the O$_1$ and O$_3$ ions in all tetrahedral units. This is illustated in Fig. \ref{Fig:OO} where the structure is shown along the \emph{pseudo} three-fold axis. In this projection the dimers are aligned perpendicular to the drawing, a dimer unit being identified by two counter-rotated tetrahedra. The OO within the dimers, not shown, is antiferro-orbital with the orbital direction as explained earlier. All the dimers in the same plane, i.e. connected by J$_2$ in the HT structure and represented by the same orbital color on the drawing, have their orbitals pointing in the same direction. However, the dimers in an adjacent plane, i.e. connected through J$_1$  (shown in different colors), have their orbitals pointing at nearly 90 $^{\circ}$ with respect to the previous layer. This unique OO pattern propagates with the same pediodicity of the optical phonon responsible for the atomic displacement, \textbf{k}=(0,0,$\frac{3}{2}$), as expected. This transition is allowed to be second order from group theory.\\
\indent The presence of OO below room temperature modifies substancially the different super-super exchange interactions between CrO$_4^{3-}$ units. This was already pointed out by Koo et al. \cite{ISI:000242899400048} who first suggested that such ordering would enhanced remarkably the strength of the intradimer interaction J$_0$, and might be required to explain the magnetic properties of Ba$_3$Cr$_2$O$_8$. In order to estimate semi-quantitatively the effect of OO on all exchange interactions, we have performed a spin-dimer analysis (\cite{CAESAR}, program freely available) based on the EHTB method (see for example \cite{ISI:000230259500004} for a review on the subject) using the crystallographic parameters at 295K and 1.6K (Table 1). Here, each Cr$_2$O$_8^{6-}$ dimer coupled along a given exchange path (J$_0$, J$_1$,...) is considered in turn, estimating the antiferromagnetic part of the exchange from the squared of the energy gap between the dimer bonding and antibonding states. This method has successfully estimated the ratios of several exchange interactions in a variety of inorganic and organic antiferromagnets including some compounds of this family \cite{ISI:000242899400048}. The calculations have been carried out using double $\xi$-Slater type orbitals, for the Cr d and O s/p orbitals, using parameters listed in the suplementary information in \cite{ISI:000242899400048}. Values of the exchange parameters, J$_i$(i=0,3), are given in Table \ref{Table:crystallo} for the HT structure. We note that J$_0$ is the dominant interaction. However, J$_1$ and J$_2$, essentially equal, are non-negligible and about 15\%\ of J$_0$. J$_3$ is negligible. This is the same trend previously reported by Koo \cite{ISI:000242899400048} for Ba$_3$Mn$_2$O$_8$ and Ba$_3$Cr$_2$O$_8$ (note that J$_1$ and J$_2$ are systematically equivalent). However, in the present study, the intra-dimer interaction is enhanced with respect to that of Ba$_3$Cr$_2$O$_8$, as expected from the shorter nearest neighbor Cr-Cr distance (3.79 compared to 3.94) due to the different chemical pressures induced by Sr and Ba. Considering the effective inter-dimer interaction parameter J' defined in the introduction, one finds a ratio J'/J$_0$=1.4 in contrast to that derived from magnetic susceptibility (J'/J$_0$=0.1 \cite{ISI:000248487900015}). In the LT structure, the three-fold rotation axis is lost, so that the interactions J$_1$ J$_2$ and J$_3$ become inequivalent along different crytallographic directions. As highlighted in Fig. \ref{Fig:OO}, the J$_1$ interactions split in a single J$_1$' (along the monoclinic a-axis) and two J$_1$'' (monoclinic b-axis). Similarly the six equivalent J$_2$ (resp. J$_3$) interactions split in two J$_2$' and four J$_2$'' (resp. two J$_3$' and four J$_3$''). The values of all exchange coupling calculated from spin-dimer analysis for the LT phase are reported in Table \ref{Table:crystallo}.  We evidence a strong increase of the intradimer interaction and a major reduction of all inter-dimer exchange with respect to the HT phase, up to a factor 50 for J$_1$'', J$_2$'. The ratio J'/J$_0$=0.11 is reduced by an order of magnitude, in perfect agreement with results obtained from susceptibility measurements estimating J$_0$ at 61.9K and J' at 6(2)K \cite{ISI:000248487900015}.
% The specific and rather unique \emph{orbital ordering} scheme appearing at high temperature is therefore the key %ingredient to promoting a network of weakly-interacting S=1/2 dimers on a three dimentional lattice, a rare %experimental realization. This is in contrast to the manganate analogues, which does not show OO probably due to %its d$^2$ electronic configuration, and retaining the trigonal symmetry down to low temperature according to Stone %and co-workers \cite{ISI:000255457200059}.
\begin{figure}[h!]
\includegraphics[scale=0.50,angle=0]{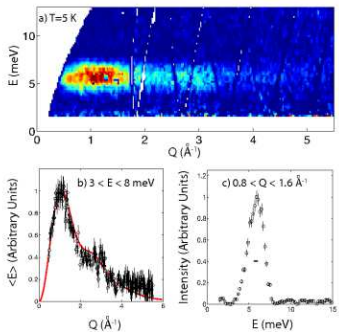}
\caption{ (Color Online) $a)$ The magnetic cross section at 5 K measured on the MARI chopper spectrometer with E$_{i}$=25 meV and a chopper frequency of 200 Hz. The first moment sum and the momentum integrated intensity are plotted in panels $b)$ and $c)$ respectively.}
\label{Fig:INS}
\end{figure}
\indent To further corroborate our results, we have investigated the inelastic magnetic response in powder samples.  The purely magnetic scattering cross-section (Fig. \ref{Fig:INS} $a)$ was obtained by considering the neutron energy gain side of the spectrum. By imposing detailed balance and assuming that the phonon cross section dominates at 175 K, a background cross section was deduced over the entire energy (E) and momentum range (Q), along with the phonon cross section, both subtracted to the total cross section, similarly to the analysis in \cite{Broholm06:74}. The resulting spectrum is illustrated in Fig. \ref{Fig:INS} $a)$ which plots the magnetic intensity as both a function of Q and E at 5 K. A large excitation gap is observed, attributed to the dimer singlet to triplet excitation, as well as a very flat band of excitations which decays with increasing Q.   
Figure \ref{Fig:INS} $b)$ plots the first moment, $\langle E \rangle$, as a function of Q, a quantity directly related to the interaction terms in the spin Hamiltonian through the Hohenberg-Brinkmann sum rule.~\cite{Hohenberg74:10}. The powder average of the first moment is written:
\begin{equation}
\langle E \rangle 
\propto|f(Q)|^2
\sum_{\bf r,d}J_{\bf d}\langle{\bf S}_{\bf r}\cdot{\bf
S}_{\bf r+d}\rangle
\left ( 1-\frac{\sin {Qd}}{Qd}\right )
\label{eq:powder}
\end{equation}
\noindent where d is the distance connecting two spins in a dimer, $f(Q)$ is the magnetic form factor, $J$ is the exchange constant and $\langle{\bf S}_{\bf r}\cdot{\bf S}_{\bf r+d}\rangle$ is the correlation function. It has been shown in many strongly dimerized systems (\onlinecite{Xu00:84}, \onlinecite{Stone02:65}) that the intra-dimer bond dominates the first moment sum. Using a single interaction term with $d=3.8 \AA$, corresponding to J$_0$, reproduces the data extremely well, as seen in Fig. \ref{Fig:INS} b).
Fig. \ref{Fig:INS} $c)$ illustrates the momentum integrated intensity as a function of E, displaying a gap at an energy of 5.46(3) meV. The full-width of 2.1 meV is substantially larger than the instrumental resolution implying a small amount of dispersion, yet difficult to determine from powder. However, the measurement does not show evidence for the strongly dispersive excitations reported for polycrystals of Ba$_3$Mn$_2$O$_8$ \cite{ISI:000255457200059}, confirming the collapse of the inter-dimer exchange coupling parameters at the orbital ordering transition. \\
\indent In summary, our neutron scattering study of the S=1/2-dimer \Sr\ compound shows that a cooperative Jahn-Teller distortion lifts the electronic degeneracy of tetrahedrally-coordinated Cr$^{5+}$ ions, inducing a dramatic reduction of the inter-dimer interactions. Small values of the inter-dimer exchange found also in Ba$_3$Cr$_2$O$_8$ with respect to the Mn compound, supports the idea that orbital ordering is universal to the chromate compounds of the series. Considering that the intra-dimer coupling energy, related to the Cr-Cr intra-dimer distance, is controllable by the chemical pressure induced by the A ions, this new class of systems appears to be ideal for finely tuning the parameters of the BEC and should become the focus of increasing attention.

%\bibliography{A3M2O8}

\end{document}